\documentclass[12pt,a4paper]{article}
\usepackage{graphicx}
\usepackage{amssymb,amsfonts,amsmath}
\usepackage{color}
\usepackage{cite}
\usepackage{bm}
\usepackage{color}

\begin{document}

\newcommand{\EQ}{Eq.~}
\newcommand{\EQS}{Eqs.~}
\newcommand{\FIG}{Fig.~}
\newcommand{\FIGS}{Figs.~}
\newcommand{\TAB}{Tab.~}
\newcommand{\TABS}{Tabs.~}
\newcommand{\SEC}{Sec.~}
\newcommand{\SECS}{Secs.~}

\title{Effects of diffusion rates on epidemic spreads in metapopulation networks}
\author{Naoki Masuda${}^{1,2}$
\\
\ \\
\ \\
${}^{1}$ 
Department of Mathematical Informatics,\\
The University of Tokyo,\\
7-3-1 Hongo, Bunkyo, Tokyo 113-8656, Japan
\ \\
${}^2$
PRESTO, Japan Science and Technology Agency,\\
4-1-8 Honcho, Kawaguchi, Saitama 332-0012, Japan\\
\ \\
masuda@mist.i.u-tokyo.ac.jp}

\setlength{\baselineskip}{0.77cm}
\maketitle

\newpage

\begin{abstract}
\setlength{\baselineskip}{0.77cm} It is often useful to represent the infectious dynamics of mobile agents by metapopulation models. In such a model, metapopulations form a static network, and individuals migrate from one metapopulation to another. It is known that heterogeneous degree distributions of metapopulation networks decrease the epidemic threshold above which epidemic spreads can occur. We investigate the combined effect of heterogeneous degree distributions and diffusion on epidemics in metapopulation networks. We show that for arbitrary heterogeneous networks, diffusion suppresses epidemics in the sense of an increase in the epidemic threshold. On the other hand, some diffusion rates are needed to elicit epidemic spreads on a global scale. As a result of these opposing effects of diffusion, epidemic spreading near the epidemic threshold is the most pronounced at an intermediate diffusion rate. The result that diffusion can suppress epidemics contrasts with that for diffusive SIS dynamics and its variants when individuals are fixed at nodes on static networks.
\end{abstract}

\newpage

\section{Introduction}\label{sec:introduction}

Infectious diseases are transmitted through social contacts between
individuals. The relationships between the structure of social networks
and the number of infected individuals
have been extensively studied
in mathematical epidemiology and network sciences.
Such studies have established that heterogeneity in contact rates enhances
epidemics
\cite{Anderson91book,Diekmann00book,Newman03siam,Barrat08book}.

The contact networks underlying, for example, sexually transmitted
diseases of humans and computer viruses on the Internet are considered to
be static on the time scale of epidemics. However, the social networks
underlying humans' prevailing infectious diseases such as influenza
are presumably dynamic even during a day. The dynamics of networks
are induced by the migration of individuals among residences, workplaces,
places for social activities, and so on.  Other animals also migrate
between habitats.  Metapopulation models are useful in describing
epidemics and ecological invasions in such a situation
\cite{Rvachev85mb,Anderson91book,Diekmann00book,Hanski00nat}.  A node in such a
model represents a metapopulation or a habitat, and not an individual.
A link represents a physical pathway connecting a pair of
metapopulations. Individuals travel from one node to another.
Simultaneously, interactions between individuals, such as infection,
can occur in each
metapopulation.

The basic reproduction number
and the condition for the occurrence of global epidemics
have been theoretically determined
for the susceptible-infected-suscepetible (SIS)
dynamics and its variants on metapopulation networks with
general connectivity profiles
\cite{Diekmann90jmb,Diekmann00book,Arino03lncis,Arino05mmb,Allen07siam}.
%
Real metapopulation networks relevant to epidemics,
such as networks of urban metapopulations and networks of airports,
often have heterogeneous degree distributions;
some metapopulations are highly connected as compared to many others
\cite{Newman03siam,Barrat08book,usair-pajek}.
Colizza et al. put important efforts for understanding
infectious dynamics on complex metapopulation networks
\cite{Colizza06pnas,Colizza07natp,Colizza07plosm,Colizza08jtb,Colizza07prl}.
In particular, they showed that
heterogeneous degree distributions enhance epidemics in uncorrelated
networks of metapopulations
(also see \cite{Fulford02tpb}). 
Similar results have been established for
models of epidemics 
on static networks of individuals
(see \cite{Newman03siam,Barrat08book}
for reviews). However,
the relationship between the epidemic threshold of the infection
rate (or population density) above which
large epidemics can occur and the degree distribution
differs in these two classes of models.

For metapopulation models, 
effects of stochasticity in infection, recovery, and migration dynamics
were also analyzed in a recent paper
\cite{Barthelemy10arxiv}.
Other authors also investigated
the threshold for infection
and epidemic 
dynamics in uncorrelated heterogeneous networks of metapopulations
\cite{Saldana08pre,Juher09pre}.

In the present work, we investigate the effect of diffusion on
epidemics in metapopulation networks on which individuals diffuse.
Diffusion increases the possibility of epidemics
for the SIS
model \cite{Jensen93jpa,Konno95jtp} and its variants
\cite{Dickman91pra,Fiore04pre} in conventional
static networks of individuals.
This is presumably
because diffusion helps infected individuals to contact new
susceptibles to be infected.
The effect of traffic-driven diffusion on the epidemic threshold
was also studied recently \cite{Meloni09pnas}.
We show that for epidemics in metapopulation networks,
diffusion prohibits rather than
enhances epidemic spreads in the sense of the epidemic threshold.
Although this result has been implicitly recognized in previous literature
\cite{Juher09pre}, we show it by developing
an analytical method to calculate the epidemic
threshold for arbitrary networks and diffusion rates.
We then support this result by numerical simulations 
and further show that an intermediate diffusion
rate magnifies epidemics near the epidemic threshold.
Qualitatively identical behavior is observed in numerical simulations
of the susceptible-infected-recovered (SIR) 
dynamics on metapopulation networks.

\section{Model}

We assume connected and undirected networks of metapopulations;
extending the
following results to the case of weighted networks is straightforward.
Each node represents a metapopulation or habitat and accommodates
any number of individuals. 
For a network having $N$ nodes,
the $N$ by $N$ adjacency matrix is denoted by $A$; $A_{ij}=1$ when
node $i$ and node $j$ are adjacent, and $A_{ij}=0$ otherwise.
$A$ is symmetric and we set $A_{ii}=0$ ($1\le i\le N$).

We consider the SIS dynamics of diffusive individuals on this network
using the continuous-time formulation developed in \cite{Saldana08pre,Juher09pre}.
The use of the original discrete-time formulation of the model
\cite{Colizza07natp} does not essentially change the following results.
The network contains $N\rho$ individuals that independently diffuse on
the network.
Each individual stays at a node and
takes either a susceptible or an infectious state.
The SIS dynamics with infection rate $\beta$ and recovery rate $\mu$
occur in each metapopulation.  
We assume mass interaction and therefore do not normalize
the infection rate by the number of individuals in the metapopulation
\cite{Colizza07natp}. By doing so, we focus on
situations in which
metapopulations that host relatively many individuals
would be a source of epidemics once an infected individual is
introduced to such a metapopulation.
The infection
event at node $i$ occurs at a rate of $\beta\rho_{{\rm S}, i}
\rho_{{\rm I}, i}$, where 
$\rho_{{\rm S}, i}$ and $\rho_{{\rm I}, i}$ are
the numbers of susceptible and infected individuals at
node $i$, respectively. 
This assumption implies all-to-all interaction
within each metapopulation. 

At the same time, individuals perform a simple random walk.
In infinitesimal time $\Delta t$, a susceptible
(infected) individual at node $i$ with degree $k_i$
moves to one of its neighboring nodes with equal probability $\approx
D_{\rm S} \Delta t/k_i$ ($\approx D_{\rm I}\Delta t/k_i$), where
$D_{\rm S}$ ($D_{\rm I}$) is the diffusion rate for the susceptible
(infected) individual and $k_i$ is the degree of node $i$.

\section{Epidemic threshold for arbitrary
  metapopulation networks}\label{sec:analysis}

We derive the epidemic threshold above which the endemic state
(\textit{i.e.}, survival of infected individuals) can occur.
Although general solutions have been formulated
\cite{Diekmann90jmb,Diekmann00book,Arino03lncis,Arino05mmb,Allen07siam}
and solutions are known for uncorrelated random networks
\cite{Colizza07natp,Saldana08pre,Juher09pre,Barthelemy10arxiv},
we are concerned with 
the effect of the diffusion rate on the epidemic threshold in general
heterogeneous networks.

The master equations are given by
\begin{align}
\frac{d\rho_{{\rm S}, i}}{dt}
&= -\beta \rho_{{\rm S}, i}\rho_{{\rm I}, i} + \mu\rho_{{\rm I}, i}
-D_{\rm S}\rho_{{\rm S}, i}+
D_{\rm S}\sum_{j=1}^N \frac{A_{ji}}{k_j}\rho_{{\rm S}, j},
\label{eq:d/dt rhoSi}\\
\frac{d\rho_{{\rm I}, i}}{dt}
&= \beta \rho_{{\rm S}, i}\rho_{{\rm I}, i} - \mu\rho_{{\rm I}, i}
-D_{\rm I}\rho_{{\rm I}, i}+
D_{\rm I}\sum_{j=1}^N \frac{A_{ji}}{k_j}\rho_{{\rm I}, j}.
\label{eq:d/dt rhoIi}
\end{align}
%
%
To derive the epidemic threshold, we consider the steady state in
which $\rho_{{\rm I}, i}$ ($1\le i\le N$) is infinitesimally small.
In this situation,
\EQ\eqref{eq:d/dt rhoSi} represents the master equation for
the continuous-time pure diffusion of susceptible individuals
because $\beta \rho_{{\rm S}, i}\rho_{{\rm I}, i}\approx 0$
and $\mu\rho_{{\rm I}, i}\approx 0$.
Given $D_{\rm S}>0$, the steady state for the number of susceptible
individuals is given by
\begin{equation}
\rho_{{\rm S}, i}^*=\frac{k_i}{\left<k\right>}\rho,
\label{eq:rhoSi in the steady state}
\end{equation}
where $\left<k\right>=\sum_{i=1}^N k_i/N$
is the average degree and
$\rho =\sum_{i=1}^N (\rho_{{\rm S}, i}+\rho_{{\rm I}, i})/N$,
the total population density.
Substituting
\EQ\eqref{eq:rhoSi in the steady state} in 
\EQ\eqref{eq:d/dt rhoIi} yields
\begin{equation}
\frac{d\rho_{{\rm I}, i}}{dt}
\approx \frac{\beta\rho}{\left<k\right>} k_i\rho_{{\rm I}, i} 
- \mu\rho_{{\rm I}, i}
-D_{\rm I}\rho_{{\rm I}, i}+
D_{\rm I}\sum_{j=1}^N \frac{A_{ji}}{k_j}\rho_{{\rm I}, j}.
\label{eq:d/dt rhoIi 2}
\end{equation}
\EQ\eqref{eq:d/dt rhoIi 2} has $\rho_{{\rm I}, i}^*=0$ ($1\le i\le N$)
as the disease-free solution. The 
destabilization of the disease-free solution
implies an endemic state, that is, $\rho_{{\rm I}, i}^*>0$
($1\le i\le N$) \cite{Diekmann90jmb,Allen07siam}.
We focus on the threshold value of
$\beta$ above which
the endemic state is induced, which is denoted by $\beta_{\rm c}$.
We note that the threshold obtained below
can also be expressed in terms of
$\rho$, as is done in previous literature
\cite{Colizza07natp,Saldana08pre,Juher09pre}, 
because $\beta$ and $\rho$ appear 
only as the multiple $\beta\rho$ in \EQ\eqref{eq:d/dt rhoIi 2}.

In the strong diffusion limit $D_{\rm I}=\infty$, 
\EQ\eqref{eq:d/dt rhoIi 2} implies
\begin{equation}
\rho_{{\rm I}, i}= \frac{k_i}{\left<k\right>}\rho_{\rm I},
\label{eq:d/dt rhoIi DB=infinity}
\end{equation}
where $\rho_{\rm I}\equiv \sum_{i=1}^N \rho_{{\rm I}, i}/N\approx 0$
is the total density of the infected individual.
By substituting \EQ\eqref{eq:d/dt rhoIi DB=infinity}
in \EQ\eqref{eq:d/dt rhoIi 2} and taking the summation over $i$, we obtain
\begin{equation}
\frac{d\rho_{\rm I}}{dt}= \left(\frac{\beta\rho\left<k^2\right>}
{\left<k\right>^2}-\mu\right)\rho_{\rm I},
\label{eq:d/dt rhoIi DB=infinity 2}
\end{equation}
which reproduces the threshold $\beta_{\rm
  c}=\mu\left<k\right>^2/\rho\left<k^2\right>$ 
known for uncorrelated networks
\cite{Colizza07natp}. 
When $D_{\rm I}=\infty$, heterogeneous degree
distributions decrease the epidemic threshold
in arbitrary metapopulation networks.

When $D_{\rm I}=0$,
\EQ\eqref{eq:d/dt rhoIi 2} represents $N$ decoupled dynamics, and
$\beta_{\rm c}$ is equal to the
epidemic threshold at an isolated node $i$ that hosts
the largest number of individuals among the $N$ nodes.
We obtain
\begin{equation}
\beta_{\rm c}= \frac{\left<k\right>\mu}
{\rho k_{\max}},
\end{equation}
where $k_{\max}$ is the degree of node $i$ \cite{Juher09pre}.
When $\beta>\beta_{\rm c}$, the nodes that satisfy
$k_i>\left<k\right>\mu/(\beta \rho)$ can eventually 
have infected individuals, whereas the other nodes can not.
In heterogeneous networks,
we obtain $\left<k^2\right>/\left<k\right>=\sum_{i=1}^N k_i^2/\sum_{j=1}^N k_j
< \sum_{i=1}^N k_ik_{\max}/\sum_{j=1}^N k_j=k_{\max}$. Therefore, 
the value of $\beta_{\rm c}$ for $D_{\rm I}=0$ is smaller than
that of $\beta_{\rm c}$ for $D_{\rm I}=\infty$.
%
%
Although this fact apparently indicates that the endemic state occurs
more easily for $D_{\rm I}=0$ than for $D_{\rm I}=\infty$, this
statement is somewhat misleading because $D_{\rm I}=0$ implies that
the infection does not travel across metapopulations.  When $D_{\rm I}=0$,
the infection is confined in the metapopulations that contain index
patients.

Assume that $D_{\rm S}$ and $D_{\rm I}$ are arbitrary.
We
develop a method to calculate the epidemic threshold for an arbitrary
structure of networks and diffusion rates.
The Jacobian $J=(J_{ij})$ of the
dynamics represented
by \EQ\eqref{eq:d/dt rhoIi 2} around the disease-free solution
is given by
\begin{equation}
J_{ij}=\left(\frac{\beta\rho k_i}{\left<k\right>}-\mu-D_{\rm I}\right)
\delta_{ij}+D_{\rm I}\frac{A_{ji}}{k_j},\quad (1\le i,j\le N)
\label{eq:Jij}
\end{equation}
where $\delta_{ij}$ is the Kronecker delta.
%
%
When the real parts of 
all the $N$ eigenvalues of $J$ are negative, the disease-free
solution is stable. Otherwise, the endemic state emerges.
Similar treatments were carried out for this model
on uncorrelated random networks
\cite{Juher09pre} and the
SIS dynamics on static networks of individuals
\cite{Boguna02pre,Boguna03prl,Wang03,Ganesh05}.

$J$ is isospectral to
\begin{equation}
J^{\prime}=(J^{\prime}_{ij})\equiv
{\rm
  diag}\left(\frac{1}{\sqrt{k_1}},\ldots,\frac{1}{\sqrt{k_N}}\right) J
{\rm diag}\left(\sqrt{k_1},\ldots,\sqrt{k_N}\right),
\end{equation}
where
${\rm diag}(d_1,\ldots, d_N)$ denotes the diagonal matrix with
the $i$th diagonal element equal to $d_i$.
Because
\begin{equation}
J^{\prime}_{ij}=
\left(\frac{\beta\rho k_i}{\left<k\right>}-\mu-D_{\rm I}\right)
\delta_{ij}
+D_{\rm I}\frac{A_{ji}}{\sqrt{k_i k_j}}
\end{equation}
is symmetric,
all the eigenvalues of $J^{\prime}$ and hence those of $J$ are
real. Let $\lambda_{\max}$ denote the maximum eigenvalue of $J$.

Fix $\beta$ and express $\lambda_{\max}$
using the Rayleigh quotient:
\begin{equation}
\lambda_{\max}=\max_{|\bm x|=1} \bm x^{\top}J^{\prime}\bm x,\quad
(\bm x\in {\mathbf R}^N)
\label{eq:Rayleigh}
\end{equation}
where $\top$ denotes the transpose. Suppose that the maximum of \EQ\eqref{eq:Rayleigh}
is realized by $\bm x = \tilde{\bm x}$, where $|\tilde{\bm x}|=1$.
When $\beta$ is replaced by $\beta + \Delta\beta$,
we obtain $\tilde{\bm x}^{\top}J^{\prime}\tilde{\bm x}=\lambda_{\max}+
\rho\Delta\beta \sum_{i=1}^N k_i \tilde{x}_i^2/\left<k\right>$, where
$\lambda_{\max}$ is the largest eigenvalue before $\beta$ is replaced by
$\beta+\Delta\beta$.
Therefore, $\lambda_{\max}$ monotonically increases with $\beta$;
this guarantees that the minimum value of
$\beta$ satisfying $\lambda_{\max}=0$ is equal to
$\beta_{\rm c}$. The condition $\lambda_{\max}=0$ is equivalent to
\begin{equation}
\det J = 0.
\label{eq:detJ=0}
\end{equation}
With regard to the value of $\beta_{\rm c}$,
\EQ\eqref{eq:detJ=0} is equivalent to
\begin{align}
0 &= \frac{\left<k\right>}{\rho}\det\left[ {\rm diag}
\left(\frac{1}{k_1},\ldots, \frac{1}{k_N}\right)J\right]\notag\\
& = \det
\left(\beta\delta_{ij}
-\frac{\mu+D_{\rm I}}{k_i}\frac{\left<k\right>}
{\rho}\delta_{ij} + D_{\rm I}\frac{\left<k\right>}{\rho}
\frac{A_{ij}}{k_i k_j}
 \right)=0.
\label{eq:rho eigenvalue}
\end{align}
Equation~\eqref{eq:rho eigenvalue} indicates that
the minimum eigenvalue of the symmetric matrix $M=(M_{ij})$, where
\begin{equation}
M_{ij}= \left(\frac{\mu+D_{\rm I}}{k_i}\delta_{ij}
-\frac{D_{\rm I}A_{ij}}{k_i k_j}
\right)\frac{\left<k\right>}{\rho},
\label{eq:rhoc final matrix}
\end{equation}
is equal to $\beta_{\rm c}$. 
For arbitrary networks, we can reliably compute the minimum eigenvalue of
$M$ in $O(N^3)$ time
by combining the Cholesky decomposition, which is applicable to
symmetric matrices, and
the inverse power method. 

For uncorrelated networks,
substituting
$A_{ij}=k_i k_j/(\left<k\right>N)$ in
\EQ\eqref{eq:rhoc final matrix} yields
\begin{equation}
M_{ij}=\left(\frac{\mu+D_{\rm I}}{k_i}\delta_{ij}
-\frac{D_{\rm I}}{\left<k\right>N}
\right)\frac{\left<k\right>}{\rho}.
\label{eq:rhoc final matrix for uncorrelated net}
\end{equation}
The inverse of
\EQ\eqref{eq:rhoc final matrix for uncorrelated net}
is given by
\begin{equation}
M^{-1}_{ij}=\frac{\rho}{\left<k\right>\left(\mu+D_{\rm I}\right)}
\left(k_i\delta_{ij} + \frac{D_{\rm I}k_ik_j}
{\mu\left<k\right>N}\right).
\end{equation}
Because all the eigenvalues of $M$ are positive, all the
eigenvalues of $M^{-1}$ are also positive. Therefore, 
we can rapidly calculate $\beta_{\rm c}$
by applying the standard power method to $M^{-1}$.
%

\section{Epidemic threshold increases with the diffusion rate}
\label{sec:beta_c vs D in theory}

In this section, we show that
$\beta_{\rm c}$ monotonically increases with $D_{\rm I}$ in arbitrary
heterogeneous networks. To this end,
we decompose $M$ as
\begin{equation}
M=(B+D_{\rm I}L)\frac{\left<k\right>}{\rho},
\end{equation}
where
\begin{equation}
B=\mu\times {\rm diag}\left(\frac{1}{k_1},\ldots, \frac{1}{k_N}\right)
\end{equation}
and
\begin{equation}
L={\rm diag}\left(\frac{1}{k_1}, \ldots, \frac{1}{k_N}\right) 
-\frac{A_{ij}}{k_ik_j}.
\end{equation}
$L$ is a Laplacian matrix 
such that its minimum eigenvalue is 0. All the other
eigenvalues of $L$ are positive for connected networks.
The interlacing eigenvalue theorem (also called Weyl's inequality)
implies that the minimum eigenvalue of the summation of two
symmetric matrices is not smaller than the sum of the minimum
eigenvalues of the two individual matrices
\cite{Hornbook,Cvetkovic10book}. 
Because the minimum
eigenvalue of $B$ is equal to $\beta_{\rm c}$ for $D_{\rm I}=0$ and
that of $L$ is equal to 0, the minimum eigenvalue of $M$,
\textit{i.e.}, $\beta_{\rm c}$ for a given $D_{\rm I}\ge 0$, is not smaller than
$\beta_{\rm c}$ for $D_{\rm I}=0$.  More generally, if $D_{\rm
I}^{\prime}\ge D_{\rm I}$, we can apply the abovementioned
arguments to the
sum of $B+D_{\rm I}L$ and $(D_{\rm
I}^{\prime}-D_{\rm I})L$, \textit{i.e.}, $B+D_{\rm I}^{\prime}L$.
Therefore, $\beta_{\rm c}$ monotonically
increases with $D_{\rm I}$.


For regular graphs, that is, networks in which all the nodes
have the same degree, $\beta_{\rm c}=\mu/\rho$ 
both for $D_{\rm I}=0$ and $D_{\rm I}=\infty$. Because $\beta_{\rm c}$
monotonically increases with $D_{\rm I}$, 
$\beta_{\rm c}=\mu/\rho$ holds true for any $D_{\rm I}\ge 0$.
The diffusion rate affects the epidemic threshold for heterogeneous
metapopulation networks but not for homogeneous networks.

\section{Numerical results}\label{sec:numerical}

The results presented
in the previous section imply that diffusion suppresses
epidemics on heterogeneous metapopulation
networks in the sense of an increased
epidemic threshold.  However, if the diffusion rate is too small,
infected individuals may be localized within the metapopulations
having index patients.  We investigate the combined effect of the
heterogeneity and the diffusion rate by carrying out direct numerical
simulations of infectious dynamics.  

We set the recovery rate
$\mu=1$ and population density $\rho=50$ without loss of generality
and vary the infection rate $\beta$. For simplicity, we set $D\equiv D_{\rm S}
= D_{\rm I}$.
%
%
We start with one infected individual selected
from the population with equal probability $1/N\rho$. The other 
$N\rho-1$ individuals are initially susceptible.
We measure the fraction of infected individuals 
in the steady state, denoted by
$\rho_{\rm I}^*$. We calculate $\rho_{\rm I}^*$
as the number of infected individuals after transient
divided by $N$ and averaged
over multiple realizations.
The relaxation time is governed by $1/D$ for small $D$ $(>0)$ and $1/\mu$
otherwise. Therefore,
we set the transient length to 500 for
$D=0$ and $D\ge 4$, and $2000/D$ for $0<D<4$.

\begin{figure}
\begin{center}
\includegraphics[width=8cm]{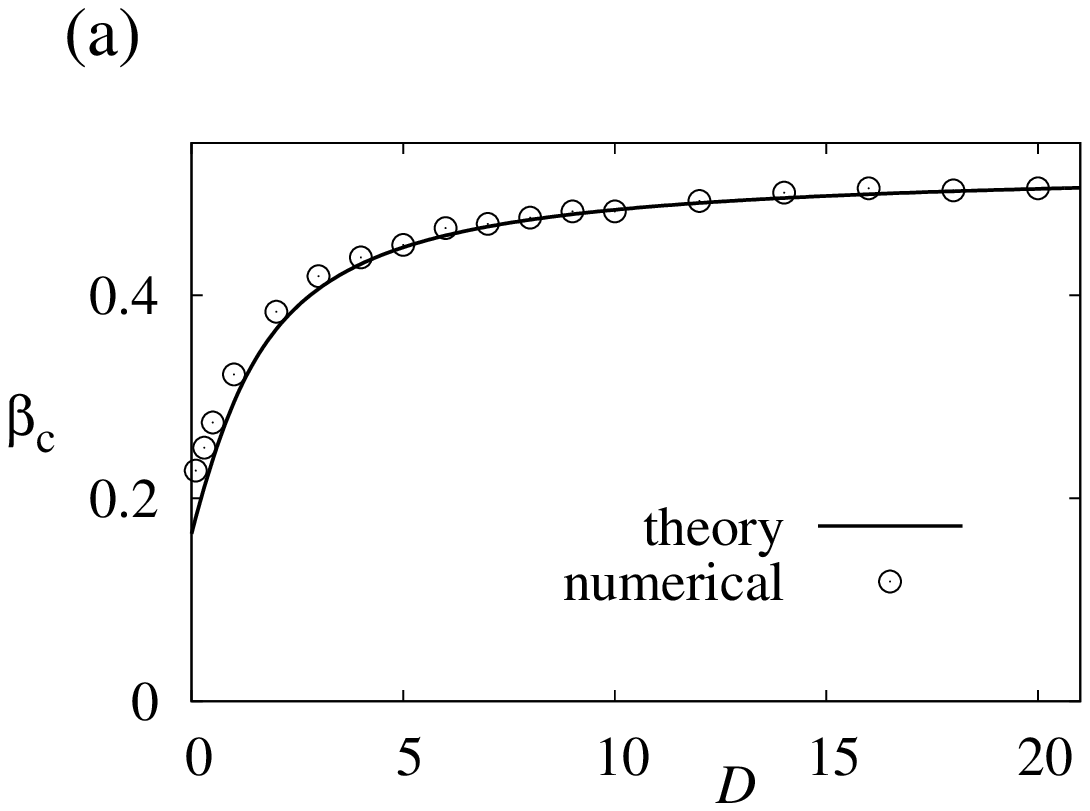}
\includegraphics[width=8cm]{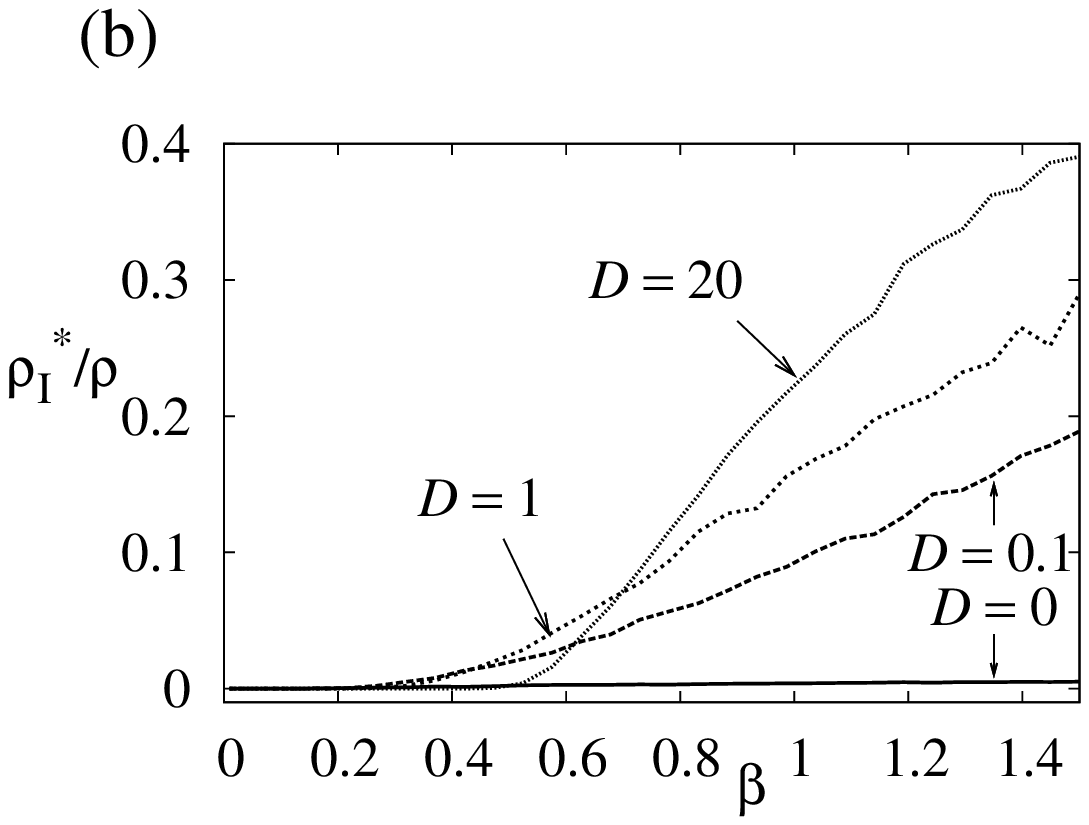}
\includegraphics[width=8cm]{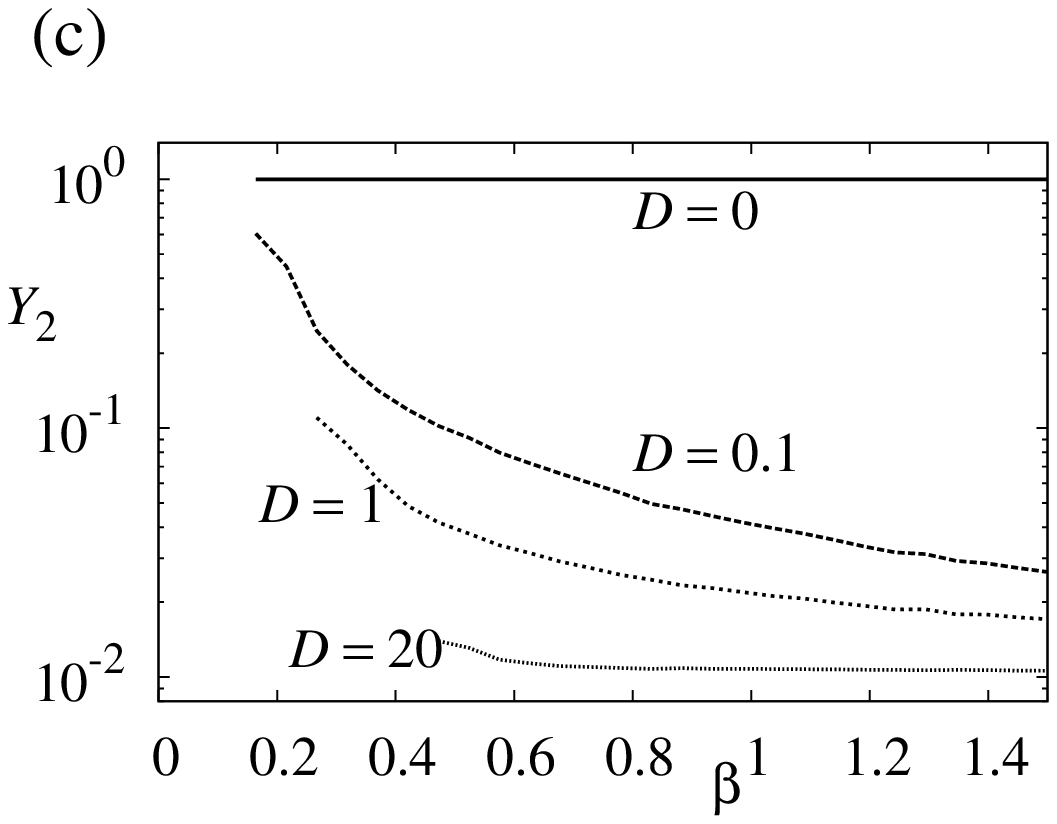}
\caption{Results for the SIS model on the scale-free metapopulation
network with $N=200$ and $\left<k\right>=6$.
(a) Relationship between $\beta_{\rm c}$ and $D$
for a generated network.
(b) Fraction of infected individuals after transient.
(c) Inverse participation ratio of the distribution of the 
infected individuals after transient.
}
\label{fig:sis}
\end{center}
\end{figure}

We first examine the SIS dynamics 
on scale-free networks generated by the Barab\'{a}si-Albert (BA) model
\cite{Barabasi99sci} with $N=200$ and mean degree 6 (\textit{i.e.}, 3
links are added per new node). To generate a network,
$N-3=197$ nodes are sequentially added to
a triangle according to the preferential attachment.
The degree distribution of the network
is the power law $p(k)\propto k^{-3}$ \cite{Barabasi99sci}.
For a generated scale-free network,
in \FIG\ref{fig:sis}(a),
we compare $\beta_{\rm c}$ obtained theoretically and numerically
for a range of $D$.
The theoretical value
of $\beta_{\rm c}$ is equal to the minimum eigenvalue of matrix $M$ given by
\EQ\eqref{eq:rhoc final matrix}.
We obtain a numerical
estimate of $\beta_{\rm c}$ as the value of $\beta$ above which an infected individual
survives in at least one of the 5000 realizations.
The numerical results are in a good agreement with the theoretical results.

Next, we examine the SIS dynamics for suprathreshold $\beta$.
For fixed $\beta$ and $D$, we generate 400 realizations 
of the scale-free network and run
the SIS dynamics 5 times.
The mean fraction of infected individuals
 $\rho_{\rm I}^*$ is calculated on the basis of 2000 runs.
The dependence of $\rho_{\rm I}^*$, normalized by $\rho$, 
on $\beta$ and $D$ are shown in \FIG\ref{fig:sis}(b).
In agreement with \FIG\ref{fig:sis}(a),
the epidemic threshold increases with the diffusion
rate. 

However, when $D=0$, $\rho_{\rm I}^*$ is small for any $\beta$. This
is because infected individuals do not migrate to other
metapopulations, and they are localized within the single
metapopulation to which the index patient belongs even above the
epidemic threshold.  When $D=0.1$, the infection reaches multiple
metapopulations above the epidemic threshold, but only to a limited
extent because of a small diffusion rate.  When $D=1$, above the
epidemic threshold, the infection seems to spread to more
metapopulations and individuals than when $D=0.1$. A
visible fraction of infected individuals ($\rho_{\rm I}^*/\rho\approx
0.03$) survives between $\beta\approx 0.4$ and $\beta\approx 0.6$, for
which larger diffusion rates do not allow the endemic state.  When
$D=20$, $\rho_{\rm I}^*$ is large for large values of $\beta$,
presumably because diffusion helps the infected individuals to reach
metapopulations with small degrees. We stress that the epidemic
threshold for $D=20$ is nonetheless larger than that for $D=0.1$ and
$D=1$, which is consistent with \FIG\ref{fig:sis}(a).

To quantify the localization
of infected individuals, in particular when
an appropriately small $D$ (\textit{i.e.}, $D=1$)
yields relatively many infected individuals
(\textit{i.e.}, $0.4\le \beta\le 0.6$),
we plot the inverse participation ratio
defined by $Y_2=\sum_{i=1}^N (\rho_{{\rm I},i}/\rho_{\rm I})^2$
(see \cite{Barthelemy04prl} for an exemplary use of $Y_2$ for
epidemic dynamics on networks).
When $Y_2$ is of the order of unity, the infected
individuals are localized in a small number of metapopulations.
When $Y_2=O(1/N)$, the infected individuals
have spread to $O(N)$ metapopulations.
In \FIG\ref{fig:sis}(c),
$Y_2$ are plotted for the values of $D$ used in \FIG\ref{fig:sis}(b).
For $D=0$, $Y_2=1$ irrespective of $\beta$, as expected.
For $D=0.1$,
$Y_2$ is relatively large for $\beta\approx 0.5$.
For $D=1$, $Y_2$ in this range of $\beta$ becomes smaller to approach
the values for $D=20$ 
(note the logarithmic scale in \FIG\ref{fig:sis}(c)).

In infectious diseases with which metapopulation models are concerned,
such as influenza,
the SIR model seems to have a wider applicability than the SIS model
\cite{Colizza06pnas,Colizza07plosm,Hufnagel04pnas}.
In the SIR model, the individuals that recover from the infected state
do not return to the susceptible state but transit to the
recovered state.
When the basic reproduction number is assumed to be the same for each
metapopulation, analytical results for the SIR model on 
heterogeneous metapopulation networks have been established
\cite{Colizza07prl,Colizza08jtb,Barthelemy10arxiv}. 
However, the analysis seems
difficult when the basic reproduction number
depends on the metapopulation. In our model, highly populated
metapopulations,
which presumably make contact with many other metapopulations,
have large basic reproduction numbers. 

\begin{figure}
\begin{center}
\includegraphics[width=8cm]{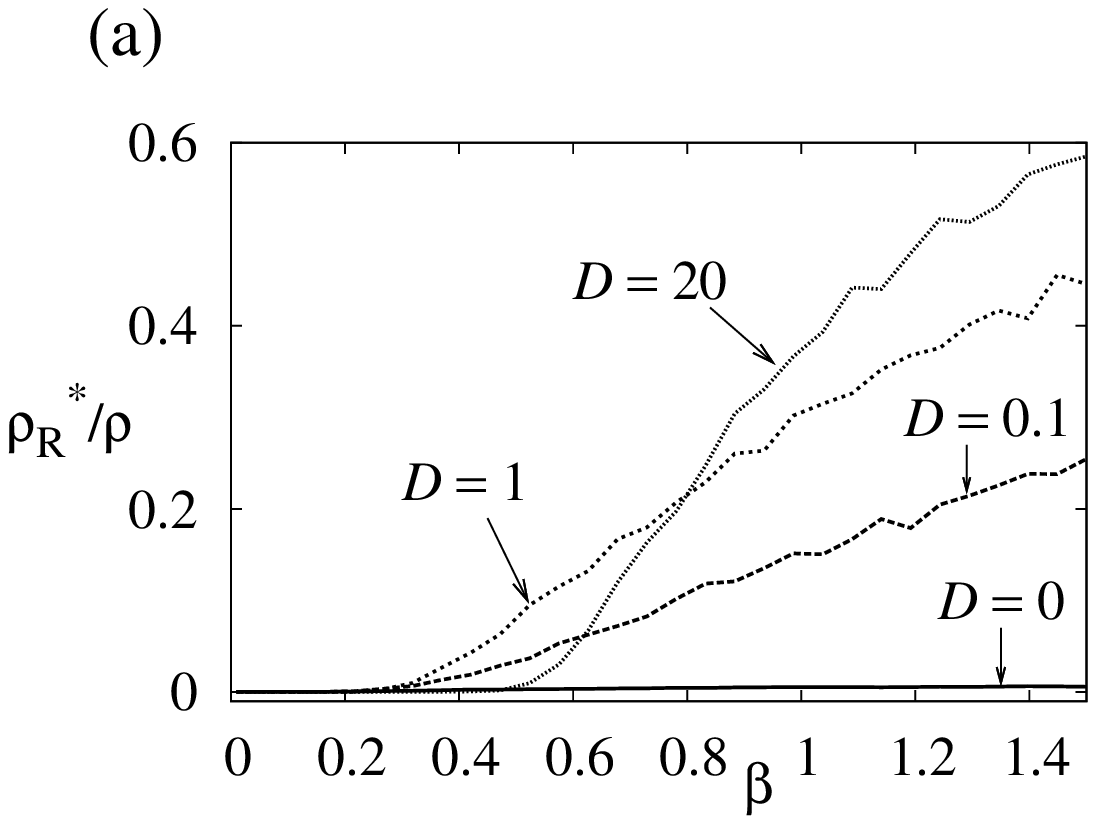}
\includegraphics[width=8cm]{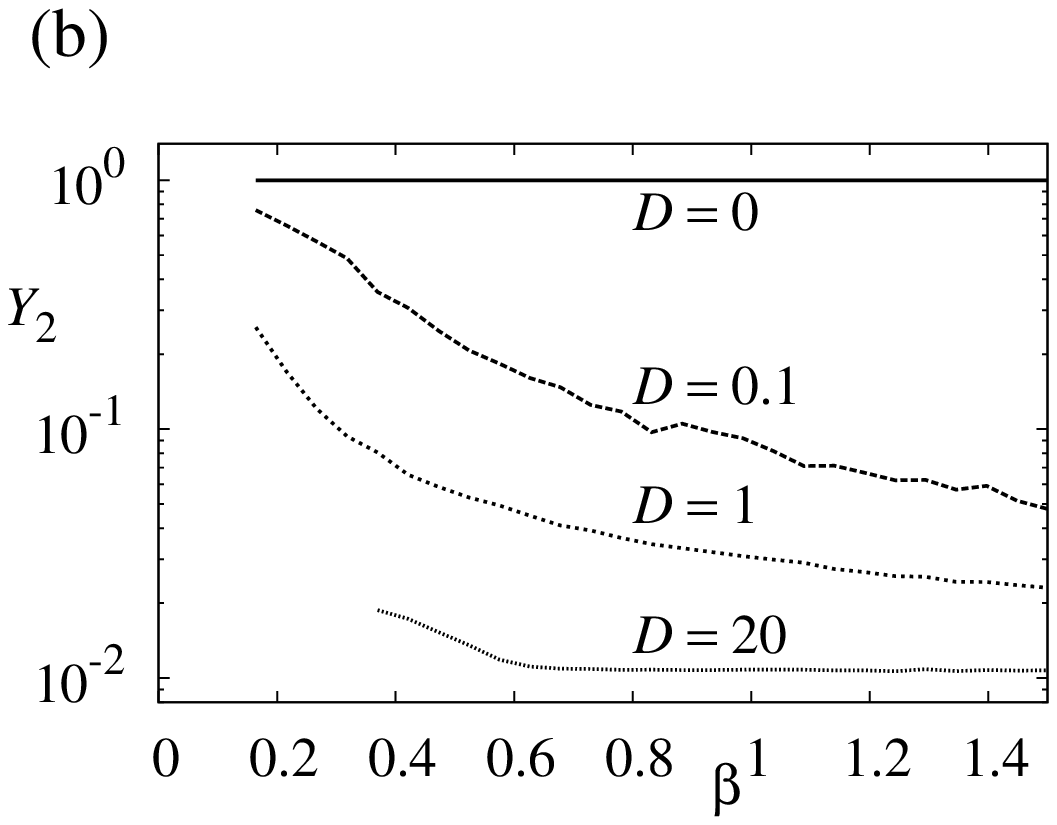}
\caption{Results for the SIR model on the scale-free metapopulation
network with $N=200$ and $\left<k\right>=6$. 
(a) Final fraction of
recovered individuals.  (b) Inverse participation ratio of the
distribution of the recovered individuals.}
\label{fig:sir}
\end{center}
\end{figure}

We carry out numerical simulations starting from an arbitrarily chosen
infected individual until there is no infected individual.
Then, we
measure the final fraction of the recovered individuals $\rho_{\rm
R}^*$ averaged over 2000 realizations. The results shown in
\FIG\ref{fig:sir}(a) are qualitatively the same as those for the SIS
model shown in \FIG\ref{fig:sis}(b).
To confirm that an appositely small $D$ facilitates spreads of
infection to many metapopulations, we measure $Y_2$. For the SIR
dynamics, we need to modify the definition of $Y_2$ because runs without
any secondary infections lead to $Y_2=1$.  Runs with a minor fraction
of infected individuals also yield a spuriously large $Y_2$.
Therefore, in the calculation of $Y_2$, we exclude the runs in which
less than 1\% of the individuals are eventually infected.  The
dependence of $Y_2$ on $\beta$ and $D$ shown in \FIG\ref{fig:sir}(b)
are qualitatively the same as those for the SIS model shown in
\FIG\ref{fig:sis}(c).

\begin{figure}
\begin{center}
\includegraphics[width=8cm]{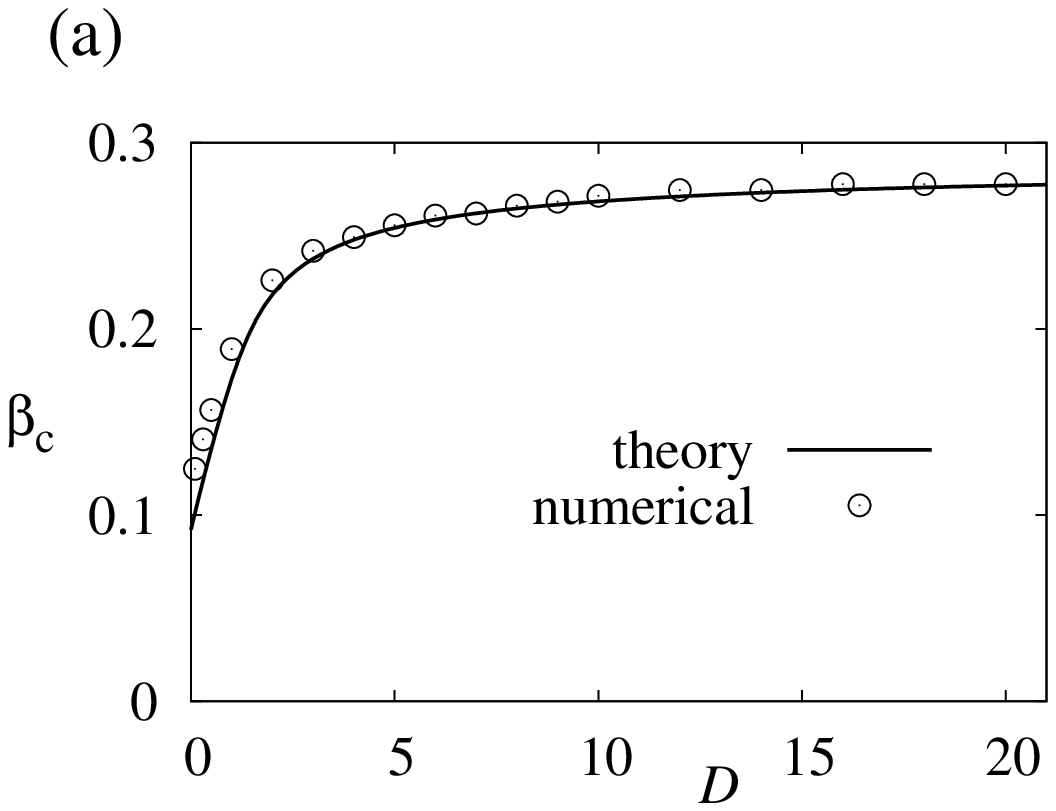}
\includegraphics[width=8cm]{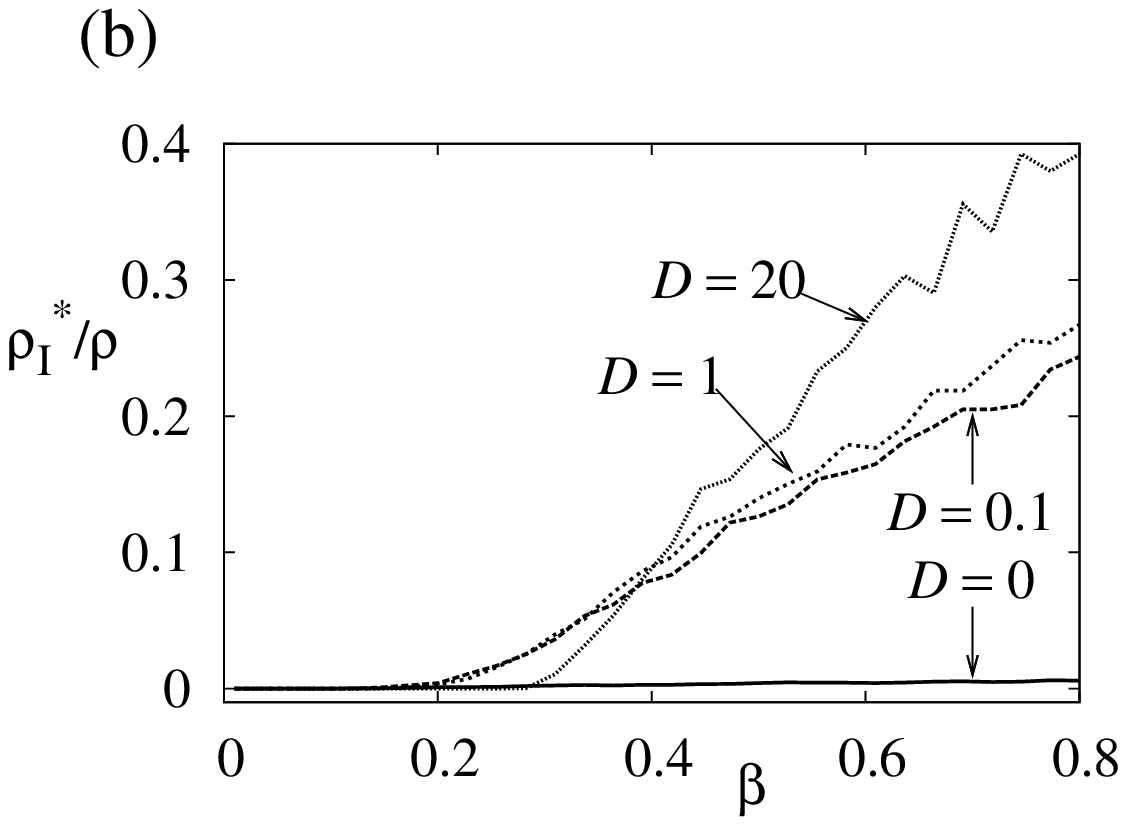}
\includegraphics[width=8cm]{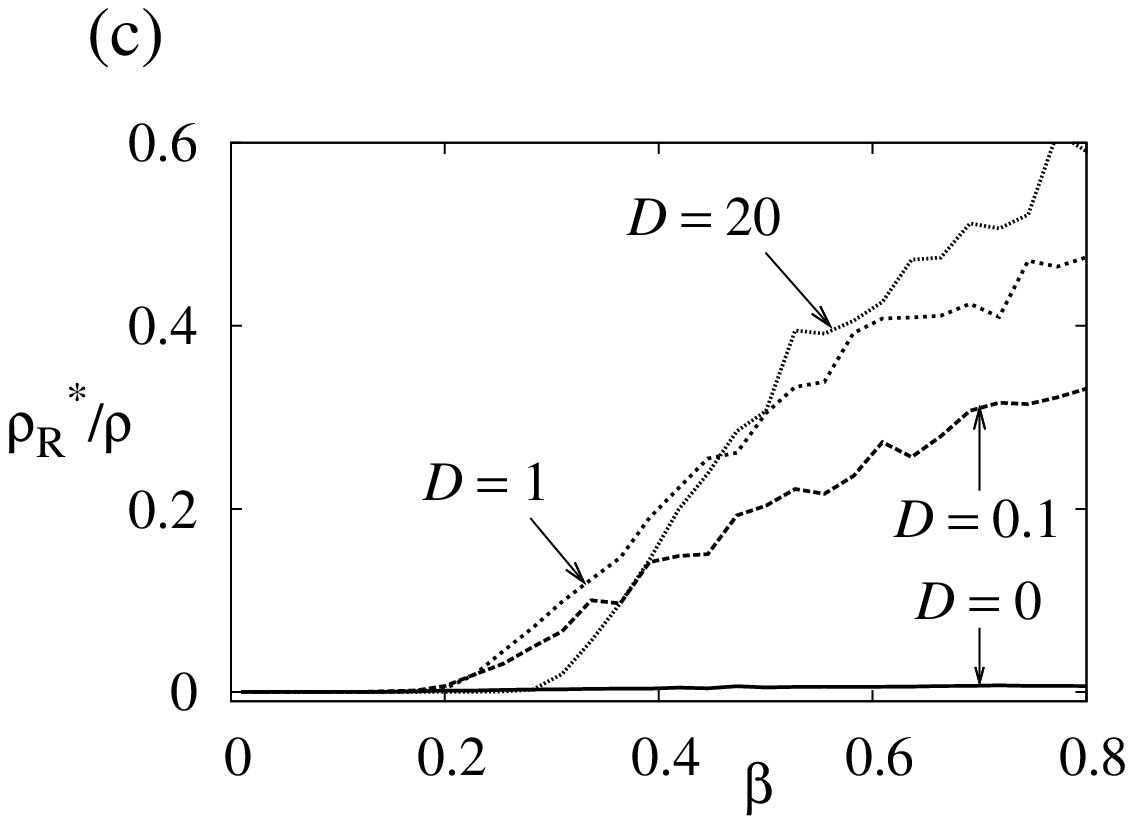}
\caption{Results for the airport network having $N=332$.
(a) Relationship between $\beta_{\rm c}$ and $D$.
(b) Fraction of
infected individuals in the SIS model after transient.  (c)
Final fraction of recovered individuals in the SIR model.
Because of the relatively large $N$, we carry out
1000 realizations for a given combination of $D$ and $\beta$ to generate
(a) and 500 realizations to generate (b) and (c).}
\label{fig:USair}
\end{center}
\end{figure}

As an example of real heterogeneous networks,
we simulate SIS and SIR dynamics
on the network of US airports
\cite{usair-pajek}. This network
has $N=332$ nodes, 2126 links, and
a long-tailed degree distribution.
We ignore the link weight.
A airport is considered to be a metapopulation that hosts traveling
individuals. 
Figure~\ref{fig:USair}(a) indicates that the values of $\beta_{\rm c}$
obtained from direct numerical simulations are in a good agreement with
the theoretical values.
The suprathreshold behavior of the SIS and SIR dynamics
shown in \FIGS\ref{fig:USair}(b) and \ref{fig:USair}(c), respectively,
is qualitatively the same as that
on the BA model (\FIGS\ref{fig:sis}(b) and \ref{fig:sir}(a)).

\begin{figure}
\begin{center}
\includegraphics[width=8cm]{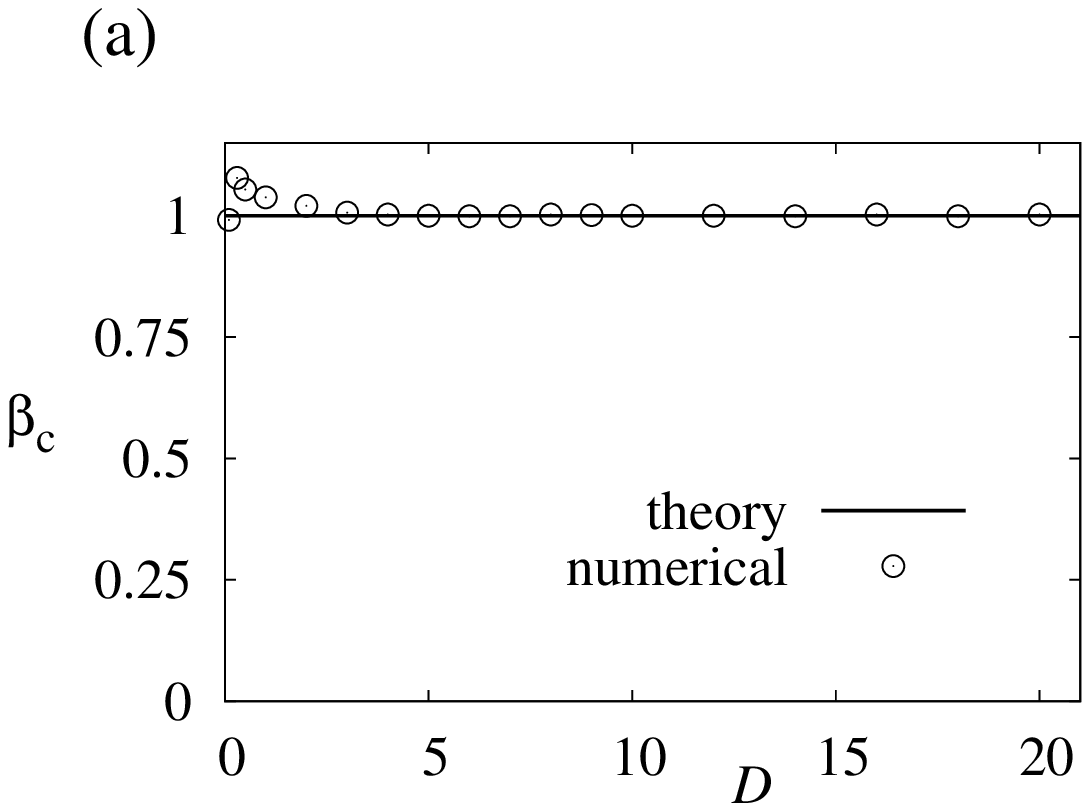}
\includegraphics[width=8cm]{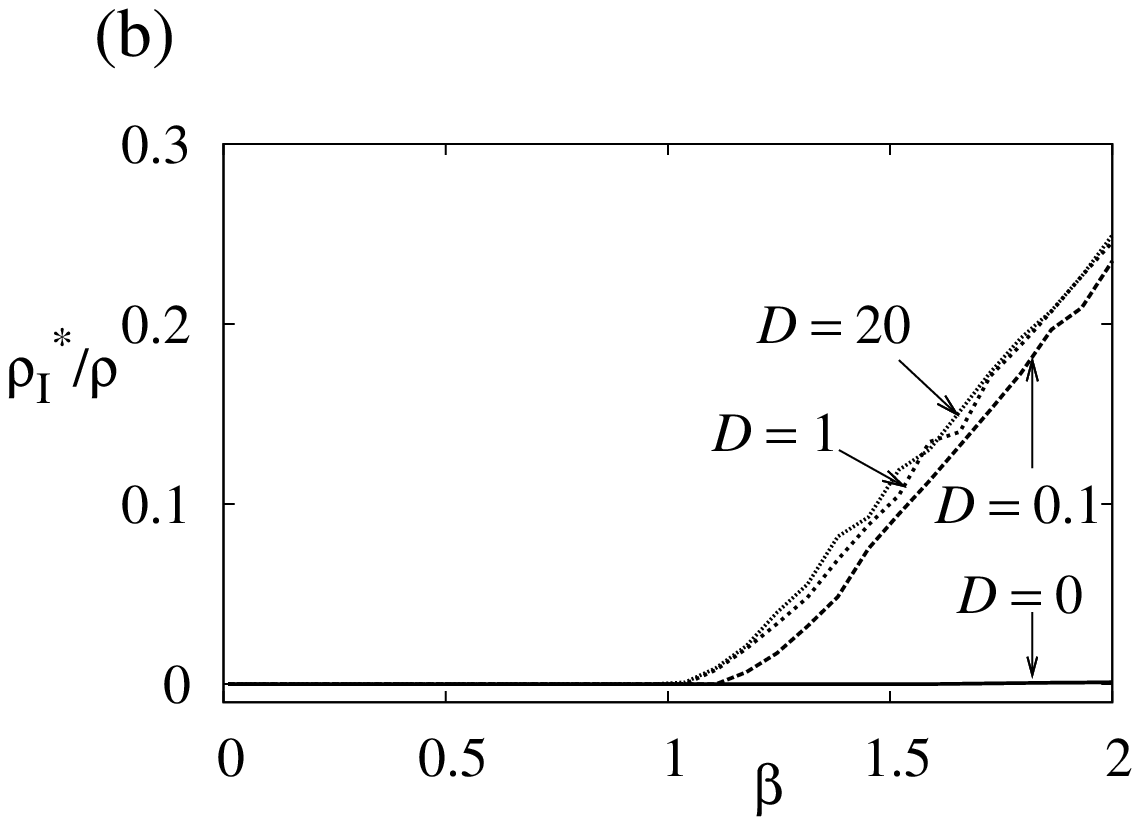}
\includegraphics[width=8cm]{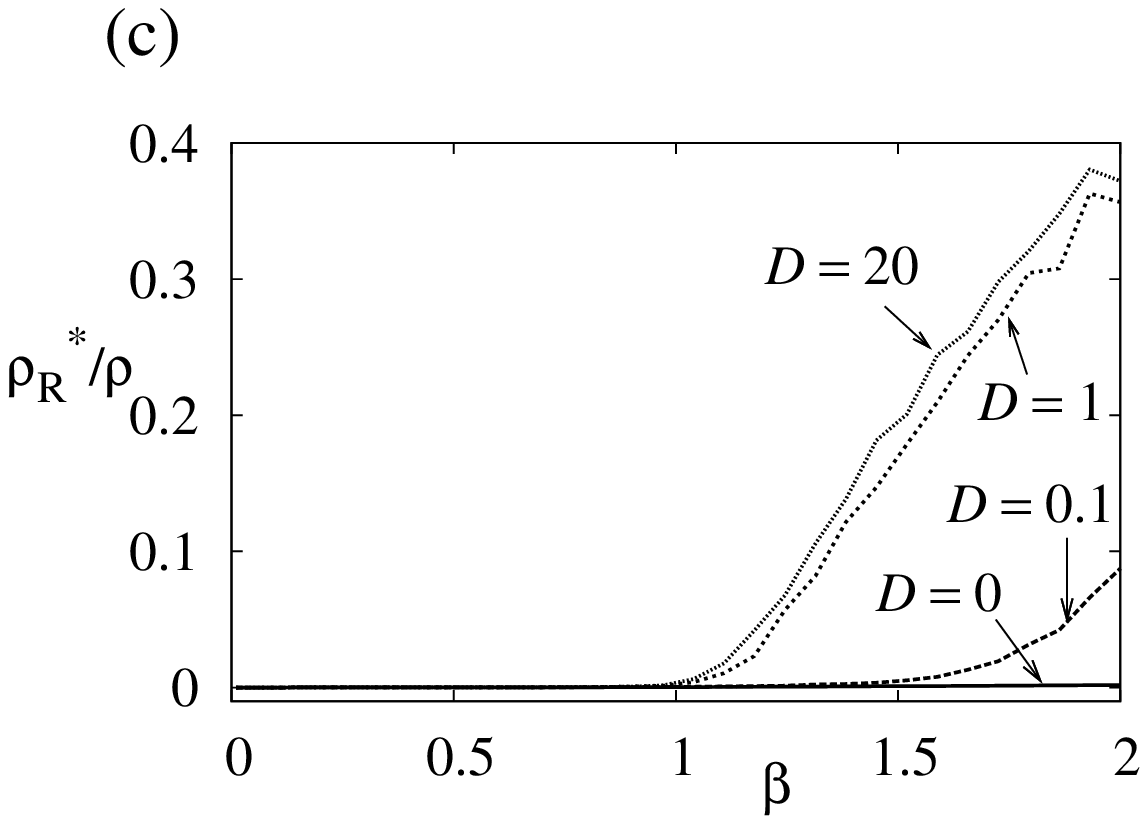}
\caption{Results for the regular random graph with $N=200$ and
$k_i=\left<k\right>=6$. (a) Relationship between $\beta_{\rm c}$ and $D$
for a generated network. (b) 
Fraction of
infected individuals in the SIS model after transient.  (c)
Final fraction of recovered individuals in the SIR model.
We carry out
5000 realizations for a given combination of $D$ and $\beta$ to generate
(a) and 2000 realizations, \textit{i.e.}, 5 realizations for each of
400 networks, to generate (b) and (c).
}
\label{fig:rrg}
\end{center}
\end{figure}

The combined effect of the heterogeneity and the diffusion rate
is absent for
homogeneous networks. To demonstrate this, we carry out
numerical simulations on the regular random graph with 
$N=200$ and degree 6;
all the nodes have degree 6 and are wired
randomly under the condition that the generated network is connected.
The theory obtained in \SEC\ref{sec:beta_c vs D in theory}
indicates that the epidemic threshold
$\beta_{\rm c}$ is independent of $D$ for the regular
random graph.
As shown in \FIG\ref{fig:rrg}(a),
the numerically obtained $\beta_{\rm c}$
is largely independent of $D$ and close to the theoretical value, 
albeit there is some disagreement for small $D$.
For a range of $\beta$, the normalized
$\rho_{\rm I}^*$ and $\rho_{\rm R}^*$ for the SIS and SIR models are shown in 
\FIGS\ref{fig:rrg}(b) and \ref{fig:rrg}(c), respectively.
Because $\beta_{\rm c}$ little depends on $D$, a large diffusion
rate results in a large fraction of infected individuals 
for any $\beta$.

\section{Discussion and Conclusions}\label{sec:discussion}

We have shown that diffusion increases the epidemic threshold of the
SIS dynamics in arbitrary heterogeneous networks.
Nevertheless, a certain amount of diffusion is
necessary to enable the transmission of epidemics across metapopulations.
Numerical simulations of the SIR dynamics also yielded similar
results. 
Similar conclusions are expected for other population dynamics
used in ecology \cite{Hanski00nat}. Indeed,
dispersal decreases the population growth
rate in ecological invasion models in which
each metapopulation carries heterogeneous birth and death
rates \cite{Schreiber09an}.
%

Our numerical results indicate that 
excessive diffusion suppresses epidemics in metapopulation networks
near the 
epidemic threshold. This effect of diffusion 
contrasts with that on the diffusive SIS model and
variants in static networks of individuals
\cite{Jensen93jpa,Konno95jtp,Dickman91pra,Fiore04pre}.
We note that, in the metapopulation model, an increased diffusion rate
can also suppress the spreading speed of epidemics.
In heterogeneous static networks of individuals,
the spreading speed in an early stage of transmission
is controlled by the largest eigenvalue of 
the matrix that represents transmission of infection
\cite{Barthelemy04prl}.
The counterpart in the metapopulation model is the matrix
given by \EQ\eqref{eq:Jij}. The largest eigenvalue of
the matrix given by \EQ\eqref{eq:Jij} decreases
with the diffusion rate; this can be shown by 
adapting the proof of the monotonic dependence of the epidemic threshold
on the diffusion rate (\SEC\ref{sec:beta_c vs D in theory}).

The nonmonotonic dependence of the fraction of infected individuals
on the diffusion rate near the epidemic threshold
is derived from the fact that
epidemics confined in a single metapopulation are irrelevant
on the global scale.
A similar distinction between global and local epidemics
is also important when analyzing 
epidemics in static networks of individuals
with community structure; epidemics confined in a single community
are not usually regarded as global epidemics \cite{Masuda09njp_immu}.


\section*{Acknowledgments}

We thank Takehisa Hasegawa for providing
valuable comments on the manuscript.
N.M. acknowledges the support through
Grants-in-Aid for Scientific Research
(Nos. 20760258 and 20540382, and
Innovative Areas ``Systems Molecular Ethology'') 
from MEXT, Japan.



\newpage
\clearpage

\clearpage

\clearpage

\clearpage

\end{document}